\documentclass{amsart}
\usepackage[latin1]{inputenc}

\usepackage{amssymb}
\usepackage{amsmath,enumerate,rotate}
\usepackage{amssymb,latexsym}
\usepackage{epsfig}
\usepackage{graphicx}
\usepackage[english]{babel}

\usepackage{a4wide}

\newtheorem{definition}{Definition}[section]

\newtheorem{lemma}[definition]{Lemma}
\newtheorem{proposition}[definition]{Proposition}

\def\CF { \mathcal{F}}
\def\CL { \mathcal{L}}
\def\CM { \mathcal{M}}

\def\CB {\mathcal{B}}

\def\CI { \mathcal{I}}

\def\CL { \mathcal{L}}

\def\CI {\mathcal{I}}
\def\CD {\mathcal{D}}

\def \CIn {\mathcal{I}}
\def\CG {\mathcal{G}}

\def\W  {{l}}  

\def\t {\tau}

\def\s {\sigma}

\def\eps {\epsilon}

\def\t {\tau}
\def\z {\zeta}

\def\J {\mathbb{I}}

\def\U {\mathbb{U}}

\def\N {\mathbb{N}}

\def\E {\mathbb{E}}
\def\RE {\mathbb{R}}

\def\setN {\mathbb{N}}

\def\T {\mathbb{T}}

\def \be {  \varpi}
\def \TT {\texttt{K}}

\usepackage{verbatim}

\begin{document}

\title{Kinetic models with randomly perturbed binary collisions}

\author{Federico Bassetti \and Lucia Ladelli \and Giuseppe Toscani }

\thanks{
Dipartimento di Matematica, Universit\`a degli Studi di Pavia, 
              via Ferrata 1, 27100, Pavia, Italy \\ 
              federico.bassetti@unipv.it}
              
\thanks{
 Dipartimento di Matematica, Politecnico di Milano,
              P.zza Leonardo da Vinci 32, 20133, Milano, Italy\\ lucia.ladelli@polimi.it}               
              
\thanks{
Dipartimento di Matematica, Universit\`a degli Studi di Pavia,                
via Ferrata 1, 27100, Pavia, Italy\\
giuseppe.toscani@unipv.it}

\begin{abstract}
We introduce a class of Kac-like kinetic equations on the real line, with general random collisional rules,  which include as particular cases
models for wealth redistribution in an agent-based market \cite{bisispigatoscani}, or  models for granular gases with a background heat bath
\cite{CarrCordTosc}. Conditions on these collisional rules which guarantee both the existence and uniqueness of equilibrium profiles and their
main properties are found.  We show that the characterization of these stationary solutions is of independent interest, since the same profiles
are shown to be solutions of different evolution problems, both in the econophysics context \cite{bisispigatoscani}, and in the kinetic theory of
rarefied gases \cite{CIS, Vil}.
\end{abstract}


\maketitle

\section{Introduction}

In this paper, we are concerned with the study of the time evolution and the asymptotic behavior of the spatially homogeneous kinetic equation
\begin{equation}
  \label{eq.boltzivp2}
  \left\{ \begin{aligned}
      & \partial_t \mu_t + \mu_t =
      {Q}^+(\mu_t,\mu_t )
       \\
      & \mu(0)=\bar \mu_0\\
    \end{aligned}
  \right.
\end{equation}
which caricatures a Boltzmann--like equation in one spatial
dimension. The solution $\mu_t =\mu_t(\cdot)$ is a time-dependent
probability measure on $\RE$, describing, in its most common
physical applications, the distribution of particle velocity in a
homogeneous gas, which is initially distributed according to the
probability measure $\bar \mu_0$. The gain operator $Q^+$ models
velocity changes due to binary particle collisions. Our fundamental
assumption is that $Q^+$ is a generalized Wild convolution. More
precisely, for all bounded and continuous test functions $g \in
C_b(\RE)$, we characterize the probability measure $Q^+(\mu ,\mu )$
by
 \begin{equation}\label{eq.Q}
  \int g(v) Q^+(\mu ,\mu)(dv)=\E \Big [\int_\RE \int_\RE
g\Big(v_1 A_1+ v_2 A_2 + A_0 \big)   \mu(dv_1) \mu(dv_2) \Big ],
 \end{equation}
where $(A_0,A_1,A_2)$  is a random vector of $\RE^3$ defined on a probability space $(\Omega,\CF,P)$ and $\E$ denotes the expectation with respect
to $P$.

The interaction rule generated by the law described in \eqref{eq.Q}
simulates an interaction in which, in addition to the standard
binary collision, the post-interaction velocities are randomly
modified by the presence of an external background. As we shall see,
this modification induces an evolution process for the probability
measure which stabilizes in time towards a steady profile heavily
dependent of this random collision part. The physical relevance of
this generalized collision rule is mainly related to the dissipative
Boltzmann equation. Indeed, in a dissipative binary collision
process, a particular choice of this random contribution is shown to
produce the same steady state of the classical Boltzmann equation
with standard dissipative binary collisions, in presence of a
thermal bath \cite{CIS}.

A second main example of application of equation \eqref{eq.Q} is
linked to the field of econophysics \cite{bisispigatoscani}. In this
case, the generalized collision refers to a market based on binary
trades between agents, in which part of the traded money is taken
away by an external third subject, which redistributes it according
to a certain economical random rule.

For $A_0=0$ and for suitable choices of $(A_1,A_2)$,
 the one-dimensional kinetic equation \eqref{eq.boltzivp2}
reduces to well-known simplified models for a spatially homogeneous
gas, in which particles move only in one spatial direction. The
basic assumption is that particles change their velocities only
because of binary collisions. When two particles collide, then their
velocities change from $v$ and $w$, respectively, to
 \begin{equation}\label{collis}
 v' = p_1 v + q_1w  \qquad w' = p_2v + q_2w
 \end{equation}
where $(p_1,q_1)$ and $(q_2,p_2)$ are two identically distributed random vectors (not necessarily independent) with the same law of $(A_1,A_2)$.

The first model of the type \eqref{eq.boltzivp2}\&\eqref{eq.Q} has
been introduced by Kac \cite{Kac}, with the collisional parameters
$p_i=\sin \tilde \theta$ and $q_i=\cos  \tilde \theta$, $i=1,2$, {
for a random angle $ \tilde \theta$, uniformly distributed  on
$[0,2\pi)$}. The dynamics describes a gas in which the colliding
molecules exchange a random fraction of their kinetic energies. This
idea has been extended in \cite{PulvirentiToscani} to gases with
inelastically colliding molecules, which loose a random part of
their energy in each interaction. The inelastic Kac equation
corresponds to \eqref{eq.boltzivp2}\&\eqref{eq.Q} with $p_i=|\sin
\tilde \theta|^p\sin \tilde\theta$ and $q_i=|\cos  \tilde
\theta|^p\cos  \tilde\theta$, with $p>0$ being the parameter of
inelasticity. Recently, more general versions of
\eqref{eq.boltzivp2}\&\eqref{eq.Q} have been considered: their
applications range from gases under the influence of a background
heat bath \cite{CarrCordTosc} to models for the redistribution of
wealth in simple market economies
\cite{CordPareTosc,MatthesToscani}. In most of the above mentioned
cases,  $A_1$ and $A_2$ are positive random variables such that
$\E[A_1^2+A_2^2]=1$ (conservation of energy) \cite{Kac,
CordPareTosc}, or $\E[A_1+A_2]=1$ (conservation of momentum)
\cite{CordPareTosc, MatthesToscani}.

In the classical Boltzmann equation \cite{Cer, Cer94} relaxation to
Maxwellian equilibrium (Gaussian density) is shown to be a universal
behavior of the solution. Contrary, the corresponding equilibria of
model \eqref{eq.boltzivp2}, to which the solution is shown to relax,
depend heavily on the precise form of the microscopic interactions
\eqref{collis}. Furthermore, they  are not always explicitly known
analytically.

In the case of models of wealth distribution in the society, the comparison of these steady states with realistic data is up to now the only means
to evaluate--- {\em a posteriori} --- the quality of a proposed model. For instance, it is commonly accepted that the wealth distribution should
approach a stationary (or, in general, a self-similar) profile for large times, and that the latter should exhibit a {\em Pareto tail}$\,\,$
\cite{review, book1}. The asymptotic behavior of the solutions of \eqref{eq.boltzivp2}, when $A_0=0$,  has been extensively treated in \cite{BaLa,BaLaMa, CeGaBo}, and it is by now fully understood, in particular if one aims to describe a few analytically accessible properties (e.g.\ moments
and smoothness).

The general situation in which $A_0 \not= 0$, while relevant in various applications which will be dealt with in this paper, has never been
touched before. This case corresponds to assume that in a binary interaction the particle velocities change from $v$ and $w$, respectively, to
 \begin{equation}\label{col-gen}
 v' = p_1 v + q_1w + \eta_1 \qquad w' = p_2v + q_2w + \eta_2
 \end{equation}
where $(p_1,q_1, \eta_1)$ and $(q_2,p_2, \eta_2)$ are two identically distributed random vectors
 with the same law
of $(A_1,A_2, A_0)$. We will now describe the specific examples we
are dealing with.


\subsubsection*{Kinetic models  of a simple market economy with redistribution}

In \cite{bisispigatoscani} Boltzmann--type kinetic models for wealth
redistribution in a simple market economy have been introduced and
discussed. The authors focused their attention to models which
include  taxes to each trading process. Assuming that a percentage
of the total wealth involved in the trade is not returned to agents,
the goal in \cite{bisispigatoscani} was to understand the role of
redistribution, there produced by a linear transport-drift type
operator.

Here we  assume that the economic trades between agents are described by an interaction of type \eqref{col-gen}. In particular,  for a given
positive constant $0 <\eps<1$, the post-interaction wealths including redistribution are given by
\begin{equation}\label{model1}
A_1=(1-\eps) \tilde A_1 \qquad
A_2=(1-\eps) \tilde A_2 \qquad
A_0=\eps \tilde A_0
\end{equation}
where $\E[\tilde A_1+ \tilde A_2]=1$ and $\E[\tilde A_0]=m_0=\int v
\bar \mu_0(dv)$. Note that within this assumption, the total mean
wealth is left unchanged.  The mixing parameters $(\tilde A_1,\tilde
A_2)$ can be chosen among the variety of models present in the
pertinent literature, see e.g. \cite{MatthesToscani}.  The classical
model introduced in \cite{ChaCha00} corresponds to  the choice
\[
\tilde A_1=\lambda +\tilde \eta (1-\lambda), \quad \tilde A_2=\tilde \eta (1-\lambda)
\]
where $\tilde \eta$ is a random variable defined on $[0,1]$ (symmetrically distributed around $1/2$)
 and $\lambda \in [0,1]$ is a parameter (the so called saving propensity), while the
pure gambling \cite{BaTo} corresponds to fix $\tilde A_1=\tilde A_2=\tilde \eta$.

An interesting variant of the previous model is obtained by setting
\begin{equation}\label{model2}
A_1=(1-\eps\Delta)\tilde A_1 \qquad
A_2=(1-\eps\Delta)\tilde A_2 \qquad
A_0=\eps \Delta \tilde A_0
\end{equation}
where $(\tilde A_1,\tilde A_2,\tilde A_0)$ and $\Delta$ are
stochastically independent,  and
$P\{\Delta=1\}=1-P\{\Delta=0\}=\delta$. The presence of $\Delta$ in
\eqref{model2} simulates a market in which taxation does not act on
the totality of trades, but it occurs only with a probability
$\delta$.

{As we shall see in Section \ref{examples-bisispigatoscani}, one can fix the values of $(A_1,A_2, A_0)$ in such a way that the steady state of the
model \eqref{eq.boltzivp2} produces the same steady states as the model considered in \cite{bisispigatoscani}.}


\subsubsection*{Inelastic Kac models with background}

{A second interesting application of binary interactions of type \eqref{col-gen} is related to the study of a dissipative gas in a thermal bath
\cite{CarrCordTosc, CIS}. In one space-dimension, a dissipative Kac-like model has been introduced and discussed in \cite{PulvirentiToscani}. As
already mentioned, this model corresponds to the choice
 \begin{equation}\label{kd}
 A_1=|\sin(\tilde \theta)|^p \sin(\tilde \theta), \quad A_2=|\cos(\tilde \theta)|^p \cos(\tilde \theta)
 \end{equation}
where $\tilde \theta$ is uniformly distributed on $[0,2\pi)$. As
shown in \cite{BaLaRe, PulvirentiToscani}, in consequence of the
dissipation, a solution to the Kac equation corresponding to an
initial value with finite second moment converges in time toward the
probability mass located in zero.  In addition to the physical
dissipative interaction \eqref{kd}, let us now assume that particles
velocities are subject to random fluctuations $\eta_i$, induced by
an external background, whose distribution is the same of $A_0$,
while $A_0$ and $(A_1,A_2)$ are stochastically independent. In
addition let us assume that $A_0\not=0$, but $\E[A_0]=0$.

{As extensively discussed in \cite{CarrCordTosc}, and directly verifiable on the single binary collision, the presence of this random fluctuation
of zero mean is such that the post-collision energy is bigger than the corresponding one 
induced by the dissipative collision without
fluctuations, i.e. when $A_0=0$. Indeed, since $A_0$ and $(A_1,A_2)$ are stochastically independent and
$\E[A_0]=0$,
 \[
 \begin{split}
\E\left( (v')^2 + (w')^2\right) & = \E(A_1^2 + A_2^2) (v^2 +w^2) + 4 \E(A_1A_2) vw + 2\E(A_0^2) \\
&= \E(A_1v+A_2w)^2+ \E(A_1w+A_2v)^2+2\E(A_0^2) \\
\end{split}
 \]
with $\E(A_0^2)>0$. The main consequence of this fact is 
that one can exhibit examples in which the initial value 
has finite second moment and at the same time the corresponding 
 solution does not converge in time toward a degenerate distribution. 
%
The same phenomenon is shown to happen if one adds to the
dissipative Boltzmann equation a thermal bath \cite{CIS}.

This allows to establish a direct link between the steady states of the present dissipative collisional models with random fluctuations and the
steady states of the dissipative Boltzmann equation in presence of diffusion \cite{CIS}, as well as in presence of friction and/or drift
\cite{Vil}. Indeed, the steady states of the various problems on the dissipative Boltzmann equation quoted above, are steady states of the
Boltzmann problem \eqref{eq.boltzivp2}, corresponding to  suitable choices of the random variables $(A_1,A_2, A_0)$.  We will detail the
correspondences between these problems in Section \ref{examples}.

}

\section{Main results}\label{main}
We start by writing the Boltzmann equation \eqref{eq.boltzivp2} in Fourier variables. By setting $\phi(t,\xi)=\int e^{i\xi v}
\mu_t(dv)$, and using Bobylev's identity \cite{Bob88}, one obtains  that $\phi(t, \xi)$ obeys to the equation
\begin{equation}
  \label{eq.boltzivp}
  \left\{ \begin{aligned}
      & \partial_t\phi(t,\xi)+\phi(t,\xi) =
      \widehat{Q}^+\Big(\phi(t,\cdot),\phi(t,\cdot)\Big)(\xi)
       \qquad(t>0, \xi \in  \RE)\\
      & \phi(0,\xi)=\phi_0(\xi) \\
    \end{aligned}
  \right.
\end{equation}
where
\begin{align}
  \label{eq.collop}
  \widehat{Q}^+ \Big(\phi(t,\cdot),\phi(t,\cdot)\Big ) (\xi)
  := \E[\phi(t,A_1\xi)\phi(t,A_2\xi) e^{i\xi A_0}].
\end{align}
The initial condition $\phi_0(\xi)=\int e^{i\xi v} \bar \mu_0(dv)$ can be seen as
 the characteristic function of a prescribed
real random variable $X_0$, i.e. $\phi_0(\xi)=\E[e^{i\xi X_0}]$.

As in the case of the Kac equation,  it is easy to see that (\ref{eq.boltzivp}) admits a unique solution $\phi$ which can be written as  a Wild
series \cite{Wild1951}
\begin{align}
\label{Wild1} \phi(t,\xi)=\sum_{n \geq 0} e^{-t}(1-e^{-t})^{n} q_n(\xi),
\end{align}
where $q_0(\xi)=\phi_0(\xi)$ and, for $n \geq 1$,
\begin{align}
\label{Wild2}
q_n(\xi)=\frac{1}{n} \sum_{j= 0}^{n-1} \hat Q^+(q_j,q_{n-1-j})(\xi).
\end{align}

Hence, if
$\mu_t$ is the unique solution of
\eqref{eq.boltzivp2} with initial condition $\bar \mu_0$, then its
Fourier-Stieltjes transform is given by \eqref{Wild1}.

\subsection{Steady states}
The stationary equation associated to  \eqref{eq.boltzivp}
is
\begin{equation}\label{stazeq}
\phi_\infty(\xi)=\hat Q^+(\phi_\infty,\phi_\infty)(\xi) \qquad (\xi \in \RE).
\end{equation}
It can be proven that, under suitable hypotheses, {a solution to
\eqref{stazeq}}  exist. To show that steady states exist it is
enough to recast the problem  as a problem of fixed point equation
for distributions. In terms of probability distributions,
\eqref{stazeq} reads
 \begin{equation}\label{stazeqVA}
Q^+(\mu,\mu) = \mu,
 \end{equation}
where, given any probability distribution $\mu$, by \eqref{eq.Q}, { the probability distribution
} $Q^+(\mu,\mu)$ is the law of the random variable
\[
A_0+Y_1A_1+Y_2A_2,
\]
$Y_1$ and $Y_2$  having law $\mu$ and $Y_1$, $Y_2$ and $(A_0,A_1,A_2)$ being stochastically independent.

In what follows, let us set
\[
\CM_{\gamma}:=\left\{ \mu \,\,\text{probability measure on} \,\, \CB(\RE):\,\, \int_\RE |x|^\gamma \mu(dx)<+\infty\right  \},
\]
and, for every $m$ in $\RE$ and $\gamma \geq 1$,
\[
\CM_{\gamma,m}:=\left \{ \mu \in \CM_\gamma \,:
 \int_\RE x \mu(dx)=m \right \}.
\]
Finally, when $\E[A_1+A_2]\not=1$, let us define
\[
\bar m:=\frac{\E[A_0]}{1-\E[A_1+A_2]}.
\]

The convex function $q:[0,\infty)\to[0,\infty]$ defined by
 \begin{equation}\label{def-q}
q(\gamma):=\E[|A_1|^\gamma+|A_2|^\gamma],
 \end{equation}
 where $0^0:=0$, will
play a very important role in what follows.

First of all,
let us  collect some known results on the existence of solutions of equation \eqref{stazeqVA}.

\begin{proposition}[\cite{rosler},\cite{ruschendorf}]\label{lemma1} Assume that there is $\gamma$ in $(0,2]$ such that
$\E[|A_0|^\gamma]<+\infty$ and
\(
q(\gamma)<1.
\)
\begin{itemize}
\item[(a)] If $0<\gamma \leq 1$, then there is a unique solution
$\mu_\infty$ of \eqref{stazeqVA} in $\CM_{\gamma}$.  In addition, if $\gamma=1$, this
solution belongs to $\CM_{1,\bar m}$;
\item[(b)] If $1<\gamma\leq 2$ and $\E[A_1+A_2]\not=1$,
then there is a unique solution
$\mu_\infty$
of \eqref{stazeqVA} in $\CM_{\gamma}$ and
this solution belongs to $\CM_{\gamma,\bar m}$;
\item[(c)] If $1<\gamma\leq 2$, $\E[A_1+A_2]=1$
and $\E[A_0]=0$, then, for every $m_0 \in \RE$, there is a unique solution
$\mu_\infty$
of \eqref{stazeqVA} in $\CM_{\gamma,m_0}$.
\end{itemize}
\end{proposition}

{ Let us notice that in case (a) (cfr. Lemma \ref{alberi-pari2}), it
is possible to describe $\mu_\infty$ in terms of a suitable series
of random variables. }

While it is easy to check when $\mu_\infty$ is a degenerate
distribution, { necessary and sufficient conditions for boundedness
of moments up to a certain order} are more difficult to obtain. A
partial answer to this problem  is given in the next proposition.

\begin{proposition}\label{momenti_beta}
 Let the same hypotheses of Proposition \ref{lemma1} be in force.
 \begin{enumerate}
\item[{\rm(i)}] { In case (a) or (b) of Proposition \ref{lemma1}
 $\mu_{\infty}$ is a degenerate distribution if and only if
$m(1-(A_1+A_2))=A_0$ almost surely {\rm(}a.s.{\rm)} for some real number $m$;
in case (c)  of Proposition \ref{lemma1}  $\mu_{\infty}$
is a degenerate distribution if and only if
$m_0(1-(A_1+A_2))=A_0$ a.s.; }
\item[{\rm(ii)}] If $q(\beta)<1$ and  $\E[|A_0|^{\beta}]<+\infty$ for
some $\beta>2$, then  $q(s) <1$ for every $\gamma \leq s \leq \beta$ and
$\int |x|^{\beta} \mu_{\infty}(dx)<+\infty $;
\item[{\rm(iii)}] { Let  $A_0$, $A_1$ and $A_2$ be positive random variables with $P\{A_0 \not=0\}>0 $.
If, for some
$\beta \geq \max \{1,\gamma\}$, $\int |x|^{\beta} \mu_{\infty}(dx)<+\infty $
and $\int x \mu_{\infty}(dx)>0$,
then $\mu_\infty\{[0,+\infty) \}=1 $ and $q(\beta)<1$.}
 \end{enumerate}
\end{proposition}

\subsection{Trend to equilibrium }
We recall that the Kantorovich-Wasserstein distance of order $\gamma>0$ between two probability measures $\mu$ and $\nu$ is defined by
\begin{align}
  \label{eq.wasserstein}
  \W_\gamma(\mu,\nu):=\inf_{(X',Y')}(\E|X'-Y'|^\gamma)^{1/\max(\gamma,1)}
\end{align}
where the infimum is taken over all pairs $(X',Y')$ of real random variables
whose marginal probability distributions  are $\mu$ and $\nu$, respectively.

If $(\nu_n)_n$ is a sequence
of probability measures belonging to
 $\CM_\gamma$ and $\nu_\infty \in \CM_\gamma$,
 then $\W_\gamma(\nu_n,\nu_\infty) \to 0$ as
 $n \to +\infty$ if and  only if
$\nu_n $ converges weakly to $\nu_\infty$ and
 \[
\int |x|^\gamma \nu_n(dx) \to \int |x|^\gamma \nu_\infty(dx).
\]
See, e.g., \cite{RachevRuschendorf}. Recall that $\nu_n $ converges weakly to $\nu_\infty$ means that $\int g(x)\nu_n(dx) \to  \int
g(x)\nu_\infty(dx)$ for every $g$ in $C_b(\RE)$.

We are now ready to state our main results concerning the long time behavior of the solutions.

\begin{proposition}\label{Pr1-Bis}
Let $\gamma \in (0,1)$. Assume that $\E[|X_0|^\gamma+|A_0|^\gamma]<+\infty$ and
\(
q(\gamma)<1
\).
Let  $\mu_\infty$ be the unique solution in $\CM_\gamma$ to \eqref{stazeqVA}.
Then,
for every $t>0$
\[
\W_\gamma(\mu_t,\mu_\infty) \leq \W_\gamma(\mu_0,\mu_\infty) e^{-t(1-q(\gamma))} .
\]
\end{proposition}

In what follows, whenever $\E|X_0|<+\infty$, set $m_0=\E[X_0]$.

\begin{proposition}\label{Pr1}
Assume that $\E[|X_0|+|A_0|]<+\infty$ and that
\(
q(1)<1
\).
Let $\mu_\infty$
be the  unique probability measure  in $\CM_{1}$ which satisfies
\eqref{stazeqVA}.  Then
for every $t>0$
\[
\W_1(\mu_t,\mu_\infty) \leq \W_1(\mu_0,\mu_\infty)e^{-t(1-q(1))}
\]
and $\int v \mu_\infty(dv)=\bar m$.
Moreover, if $m_0=\bar m$, then $\int v \mu_t(dv)=\bar m$ for all $t \geq 0$.
\end{proposition}

\begin{proposition}\label{Pr2}
Assume that, for some $\gamma \in (1,2]$, $\E[|X_0|^{\gamma}+|A_0|^{\gamma}]<+\infty$,
\(
q(\gamma)<1,
\)
$\E[A_1+A_2]\not=1$ and $m_0=\bar m$.
Let  $\mu_\infty$ be the unique solution in $\CM_\gamma$ to \eqref{stazeqVA}.
Then, for every $t >0$,
\[
\W_\gamma(\mu_t,\mu_\infty) \leq 2^{1/\gamma} \W_\gamma(\mu_0,\mu_\infty)e^{-t(1-q(\gamma))/\gamma}
\]
and $\int_\RE v \mu_\infty(dv)=\bar m$.
Moreover, $\int v \mu_t(dv)=\bar m$ for all $t \geq 0$.
\end{proposition}

\begin{proposition}\label{Pr3}
Assume that,  for some $\gamma \in (1,2]$, $\E[|X_0|^{\gamma}+|A_0|^{\gamma}]<+\infty$,
\(
q(\gamma)<1
\)
and that $\E[A_1+A_2]=1$ and $\E[A_0]=0$.
Let $\mu_\infty$ be the unique solution in $\CM_{\gamma,m_0}$ of \eqref{stazeqVA}.
Then
\[
\W_\gamma(\mu_t,\mu_\infty) \leq 2^{1/\gamma}\W_\gamma(\mu_0,\mu_\infty)e^{-t(1-q(\gamma))/\gamma}
\]
for every $t >0$. Moreover, $\int v \mu_t(dv)=m_0$ for all $t \geq 0$.
\end{proposition}

\section{Examples}\label{examples}

\subsection{Kinetic models  of a simple market economy with redistribution}
The first application of the results of Section \ref{main} deals with the kinetic model for wealth with redistribution,  briefly described in the
Introduction. In this leading example, the random variables $A_1, A_2, A_0$ are given by  \eqref{model2}.
As already noticed, these assumptions correspond to a kinetic model for wealth distribution in which part of the wealth put into the binary trade is taken away by a
third subject, which at the same time restitutes to agents a certain amount of wealth. This is done
in such a way that the mean total amount of wealth into the
system is left unchanged.
Assuming \eqref{model2}, one has
\[
\E[A_1+A_2]=1-\eps\delta < 1.
\]
Hence, since $\E[\tilde A_0]=\int v \bar \mu_0(dv)=m_0<+\infty$, one
can invoke  Propositions \ref{lemma1} (a) and  \ref{Pr1} to prove
both the existence and uniqueness in $\CM_1$ of a steady state and the
(exponential) convergence to this steady state of any solution with
finite initial moment of order one.

{ One of the interesting effect of the redistribution is that the
steady state  can have finite moments of higher order than those of
the steady states of the corresponding model without
redistribution.} This can be easily verified by comparing the steady
states corresponding to $\eps>0$, say
$\mu_{\infty}^{(\delta,\eps)}$, with the steady states without
redistribution, we will denote by  $\mu_{\infty}^{(0,0)}$, obtained
by setting $\delta=\eps=0$. Thanks to Theorem 5.3 in
\cite{DurretLiggett}, it is  known that, given $\beta>1$,
 $\int v^\beta \mu_{\infty}^{(0,0)}(dv) <+\infty $ if and only if
 $\tilde q(\beta):=\E[\tilde A_1^\beta+\tilde A_2^\beta]<1$.
On the other hand, if $\eps>0$ and $\E[A_0^{\beta}]<+\infty$, by
Proposition \ref{momenti_beta}  (ii)-(iii), $\int v^\beta
\mu_\infty^{(\delta,\eps)}(dv) <+\infty $ if and only if
$q(\beta)<1$. Since $q(\beta)=[1+\delta[(1-\eps)^\beta-1]]\tilde
q(\beta)$ and $[1+\delta[(1-\eps)^\beta-1]]<1$, one can easily give
examples in which $\int v^\beta \mu_{\infty}^{(0,0)}(dv) =+\infty $
while $\int v^\beta \mu_\infty^{(\delta,\eps)}(dv) <+\infty $.

The previous discussion does not solve another interesting problem
connected with wealth taxation and redistribution:  the existence of
an \emph{optimal} amount of taxation. If one assumes that, given a
certain (conserved) amount of money, the optimal redistribution
refers to a steady state in which all people in the market ends up
with almost the same amount of money, this problem can be solved by
looking for the steady state with minimal variance. We leave this
point to a further research.

\subsection{Connections
with other form of redistribution}\label{examples-bisispigatoscani}

We show here that the law of $(A_1,A_2, A_0)$ can be fixed in such a way that the steady states of the redistribution model proposed in
\cite{bisispigatoscani} fit into our framework.

Let us start by briefly outlining the model introduced in
\cite{bisispigatoscani}. In Fourier variables this model reads
\begin{equation}\label{eqbisispiga}
\frac{\partial}{\partial t} \phi(t,\xi) + \phi(t,\xi) =
      \hat Q_\eps(\phi,\phi)(t,\xi) + \hat R_\chi^\eps(\phi)(t,\xi)
\end{equation}
with
 \begin{equation}\label{coll1}
\hat Q_\eps(\phi,\phi):=\E\big [ \phi \big ( A_1^* \xi \big)\phi
\big ( A_2^*\xi \big)],
 \end{equation}
  and
 \begin{equation}\label{redis}
\hat R_\chi^\eps(\phi)(\xi):=-\eps \chi \xi \frac{\partial }{\partial \xi} \phi(\xi)
+ i \eps (\chi+1)m_0 \xi \phi(\xi) \qquad (\chi \geq -1),
 \end{equation}
In \eqref{coll1}  $(A_1^*, A_2^*)$ are positive random variables such
that $\E[ A_1^*+\ A_2^*]=1-\eps$,  and {$\frac{\partial }{\partial
\xi} \phi(0,0)=i\int v \bar \mu_0(dv)=i m_0 $.
 Note  that in \eqref{eqbisispiga} the
interaction operator consists  in a \emph{dissipative} collision
operator, given by $\hat Q_\eps(\phi,\phi)$, and a redistribution
(differential) operator $\hat R_\chi^\eps(\phi)(\xi)$.  It is worth
recalling that, if $\phi$ is the Fourier-Stieltjes transform of  a
(regular) density $f$, then  $\hat R_\chi^\eps(\phi)$ is the
Fourier-Stieltjes transform of
\[
R_\chi^\eps(f)(v)=\eps \frac\partial{\partial v} \Big[ \left(\chi v - (\chi + 1)m_0\right)f(v) \Big].
\]
}
 The possible steady states of \eqref{eqbisispiga} must satisfy
\begin{equation}\label{staztoscanibisispiga}
\phi(\xi) =
      \hat Q_\eps(\phi,\phi)(\xi) + \hat R_\chi^\eps(\phi)(\xi).
\end{equation}
Existence of a global solution  $\phi(\xi,t)$ to \eqref{eqbisispiga}
has been proved in \cite{bisispigatoscani} provided that $\int v\bar
\mu_0(dv)=m_0$.  Anything was proven about the existence (and
eventually uniqueness) of a steady state.
This problem can be solved in
a surprisingly easy way by establishing a connection between the steady states of the model \cite{bisispigatoscani} and  special cases of our model.

First of all let us fix $\chi=-1$ in \eqref{redis}. In this case the
redistribution operator simplifies, and equation
\eqref{staztoscanibisispiga} reduces to
\begin{equation}\label{aaa}
\phi(\xi)= \hat Q_\eps(\phi,\phi)(\xi ) +\eps \xi \frac{\partial }{\partial \xi} \phi(\xi).
\end{equation}
Resorting to the analogous computation in Bobylev, Cercignani and
Gamba \cite{BoCe,CeGaBo}, equality \eqref{aaa} can be equivalently
rewritten as
 \begin{equation}\label{drift2}
 \phi(\xi)=\int_0^1  \hat Q_\eps(\phi,\phi)(\xi u^{-\eps})du.
 \end{equation}
It is immediate to see that equation \eqref{drift2} can be rephrased
as
 \begin{equation}\label{modelloselfsimil}
\phi(\xi)=\E\big [ \phi \big ( U^{-\eps} A_1^* \xi \big)\phi \big ( U^{-\eps} A_2^*\xi \big)],
\end{equation}
where $U$  and $(A_1^*,A_2^*)$ are stochastically independent and
$U$ is uniformly distributed on $[0,1]$. Hence, the steady state
\eqref{aaa} coincides with the steady state \eqref{stazeq} corresponding
to $(A_0,A_1,A_2):=(0,U^{-\eps} A_1^*,U^{-\eps} A_2^*)$. Since in
this case $q(1)=\E[A_1+A_2]=1$, in order to apply Proposition
\ref{lemma1} (c) it is necessary that $A^*_1$ and $A^*_2$ satisfy
\[
q(\gamma)=\frac{1}{1-\gamma\eps} \E[(A_1^*)^\gamma+(A_2^*)^\gamma]<1,
\]
for some $1<\gamma\leq \min\{2,1/\eps \}$. If
$\E[(A_1^*)^\gamma+(A_2^*)^\gamma]<1$ this inequality { holds true
for every $\eps<(1-\E[(A_1^*)^\gamma+(A_2^*)^\gamma])/\gamma $.} It
should be noticed that in this case, since $A_0=0$, and $A_1$ and
$A_2$ are positive with $\E[A_1+A_2]=1$ one can resort also to
Theorem 2(a) of \cite{DurretLiggett}.

Let us now consider the case in which $\chi=0$, so that
\eqref{redis} corresponds to a pure transport operator, {which
produces a uniform redistribution.}  In this case,  equation
\eqref{staztoscanibisispiga} becomes
  \begin{equation}\label{staz2}
\phi(\xi) =  \hat Q_\eps(\phi,\phi)(\xi) + i\eps m_0 \xi \phi(\xi),
 \end{equation}
or, what is the same,
\begin{equation}\label{transportcase}
\phi(\xi) =  \frac{1}{1-i\eps m_0\xi} \hat Q_\eps(\phi,\phi)(\xi).
\end{equation}
Let us observe that if $A_0$ is an exponential random variable of mean $\eps m_0$, that is with density $h_0(v)=\exp\{-v/(\eps m_0)\}/({\eps m_0})$
($v
>0$), then
\[
\E[e^{i\xi A_0}]=\int_0^{+\infty} e^{i\xi v} h_0(v)dv=\frac{1}{1-i\eps m_0 \xi}.
\]
Under the additional assumption that $A_0$ and $(A_1^*,A_2^*)$ are stochastically independent, \eqref{transportcase} can be equivalently written
as
\[
\phi(\xi)=\E\big [ e^{i\xi A_0} \phi \big ( A_1^* \xi \big)\phi \big ( A_2^*\xi \big)].
\]
Hence, it is enough to choose $A_1=A_1^*,A_2=A_2^*$ and $A_0$ as
above to identify the steady state \eqref{staz2}  with the steady
state \eqref{stazeq}. Note that, since in this case
$\E[A_1+A_2]=1-\eps<1$, the assumptions of Proposition \ref{lemma1}
(a) are trivially satisfied for $\gamma=1$ .

Last, let us examine the physically relevant case in which $\chi>-1$
and $\chi\not =0$.  { For any given $\eps \in (0,1]$ set
$\delta:=\eps \chi>-\eps$. With this choice,
\eqref{staztoscanibisispiga} becomes
\[
\hat Q_\eps(\phi,\phi)(\xi)=\phi(\xi)+\delta\xi \frac{\partial }{\partial \xi} \phi(\xi)-i(\delta+\eps)m_0 \xi \phi(\xi).
\]
Multiplying both sides for $e^{-i\xi m_0 \frac{\delta+\eps}{\delta}}\xi^{\frac{1}{\delta}-1}$
we get
\[
\begin{split}
\hat Q_\eps(\phi,\phi)(\xi) e^{-i \xi m_0 \frac{\delta+\eps}{\delta} } \xi^{\frac{1}{\delta}-1}
& =
e^{-i\xi m_0 \frac{\delta+\eps}{\delta}}\xi^{\frac{1}{\delta}-1}
\Big (\phi(\xi)+\delta\xi \frac{\partial }{\partial \xi} \phi(\xi)-i(\delta+\eps)m_0 \xi \phi(\xi)\Big)\\
&=
\frac{\partial}{\partial \xi}\Big ( \delta e^{-i \xi m_0 \frac{\delta+\eps}{\delta}} \xi^{\frac{1}{\delta}}
\phi(\xi)\Big),\\
\end{split}
\]
which, integrating over $[0,\xi]$, gives
 \begin{equation}\label{gen2}
\delta e^{-i \xi m_0 \frac{\delta+\eps}{\delta}} \xi^{\frac{1}{\delta}}
\phi(\xi)=\int_0^\xi \hat Q_\eps(\phi,\phi)(\tau) e^{-i \tau m_0 \frac{\delta+\eps}{\delta} } \tau^{\frac{1}{\delta}-1} d\tau.
 \end{equation}
By the change of variable $\xi u^\delta=\tau$, we can write the previous equation in the equivalent form
\[
\phi(\xi)=\int_0^1 \hat Q_\eps(\phi,\phi)(\xi u^\delta) e^{i(1-u^\delta) \frac{\delta+\eps}{\delta} m_0\xi} du,
\]
which can be rephrased as
\[
\phi(\xi)=\E[\phi(U^\delta  A_1^*\xi)\phi(U^\delta
A_2^*\xi)e^{i(1-U^\delta) \frac{\delta+\eps}{\delta} m_0\xi}],
\]
where $( A_1^* , A_2^* )$ and $U$ are stochastically independent and $U$ is uniformly distributed on $[0,1]$.
Hence  \eqref{staztoscanibisispiga} is equivalent to
\[
\phi(\xi)=\E[\phi(A_1 \xi)\phi(A_2 \xi)e^{i A_0\xi}]
\]
for
\begin{equation}\label{gen1}
 A_1=U^{\delta}  A_1^* \qquad A_2=U^{\delta}  A_2^*  \qquad A_0=(1-U^{\delta}) \frac{\delta+\eps}{\delta}m_0.
 \end{equation}
Since $\delta>-\eps$, it is immediate to reckon that
$\E[A_1+A_2]=(1-\eps)/(1+\delta)<1$. Hence Proposition \ref{lemma1}
(a) applies. In particular, since $\E[A_0]/(1-\E[A_1+A_2])=m_0$, the
solution $\mu_\infty$ described in Proposition \ref{lemma1}
satisfies $\int v \mu_\infty(dv)=m_0$.}

\subsection{Inelastic Kac models with background and connection with dissipative models with diffusion}

A further application of the results of Section \ref{main},  announced in the Introduction, results from the choice
 \begin{equation}\label{ka1}
 A_1=|\sin(\tilde
\theta)|^p \sin(\tilde \theta), \quad A_2=|\cos(\tilde \theta)|^p \cos(\tilde \theta)
 \end{equation}
with $\tilde \theta$ uniformly distributed on $[0,2\pi)$. This
assumption leads, when $A_0=0$, to the inelastic Kac model
\cite{PulvirentiToscani}, which describes the cooling of a
one-dimensional  spatially homogeneous Maxwell--like gas. In fact,
if $A_0=0$ and $\int |v|^{2/(1+p)} \mu_0(dv)<+\infty$,  $\mu_t$ is
shown to converge weakly to the probability mass concentrated in $0$
(cfr. \cite{BaLaRe}). As we shall see, the addition of a random
fluctuation, described by the random variable $A_0\not=0$, is
responsible for the formation of non-trivial steady states.  If $A_1$
and $A_2$ are given by \eqref{ka1}, then
\[
\E[A_1+A_2]=0,
\]
and, whenever $\gamma>2/(1+p)$,
\[
q(\gamma)=\E[|A_1|^\gamma+|A_2|^\gamma] =\frac{1}{2\pi} \int_0^{2\pi}[ |\sin(\theta)|^{(p+1)\gamma}+|\cos(\theta)|^{(p+1)\gamma} ]d\theta<1.
\]
{ Hence, if for some positive $\eps$,  with $\eps +2/(1+p)<2$,
$\E|A_0|^{2/(1+p)+\eps}<+\infty$ (with $\E[A_0]=m_0:=\int v
\mu_0(dv)$ if $p<1$) and
 $\int |v|^{2/(1+p)+\eps}  \bar \mu_0(dv)<+\infty$, then Propositions \ref{Pr1-Bis}-\ref{Pr2} apply.} In particular,
 if  $\E[A_0]=0$ and $P\{A_0 \not = 0\}>0$,
 the steady state is a non-degenerate probability distribution with finite moments of all orders.

As a special case let us choose $A_0=A_{0,a}-A_{0,b}$
with $A_{0,a}$  and $A_{0,b}$
exponentially distributed with density \( v\mapsto \exp\{-{v}/{a}\}/{a} \) and
 \( v\mapsto \exp\{-{v}/{b}\}/{b} \) $(v>0)$, and assume that $A_{0,a},A_{0,b},A_1,A_2$ are stochastically
  independent. Since
\[
\E[e^{i\xi A_0}]=\E[e^{i\xi A_{0,a}}]\E[e^{-i\xi A_{0,b}}]=\frac{1}{(1-ia\xi)(1+ib\xi)}
=\frac{1}{1-i(a-b)\xi+ab\xi^2},
\]
if $a:=(m_0+\sqrt{m_0^2+4\sigma^2})/2$ and $b:=(-m_0+\sqrt{m_0^2+4\sigma^2})/2$
the stationary equation \eqref{stazeq} becomes
\[
\E[\phi(A_1\xi)\phi(A_2\xi)]=\phi(\xi)(1-i m_0\xi+\sigma^2\xi^2)
\]
which can be equivalently written, after setting  $\hat Q_p^+(\phi,\phi) := \E[\phi(A_1\xi)\phi(A_2\xi)]$, as
 \begin{equation}\label{diff}
\phi(\xi)= \hat Q_p^+(\phi,\phi)(\xi)  -\sigma^2 \xi^2 \phi(\xi) +i m_0  \xi \phi(\xi).
 \end{equation}
{ Equation \eqref{diff} describes the steady states of the inelastic
Kac equation in presence of a thermal bath and a transport term.
Indeed, if $\phi$ is the Fourier-Stieltjes transform of a density
$f$, then $-\sigma^2 \xi^2 \phi(\xi) +i m_0  \xi \phi(\xi)$ is the
Fourier-Stieltjes transform of
\[
\sigma^2 \frac{\partial^2}{\partial v^2}
f(v,t) - m_0 \frac{\partial}{\partial v} f(v,t).
\]
}In particular, the analysis of Section \ref{main} allows to prove existence of a steady state for the dissipative Kac equation with diffusion. The
problem of the solvability of equations of type
 \begin{equation}\label{hb}
 Q(f,f) + \sigma^2  \Delta f = 0,
 \end{equation}
 in terms of nonnegative integrable densities $f \in L^1_+(\RE^3)$, and where $Q$ is the Boltzmann collision operator, is a well-known problem in
 kinetic theory of rarefied gases. When $Q$ is the dissipative collision operator for Maxwellian molecules, existence of non trivial weak
 solutions has been proved by Cercignani, Illner and Stoica \cite{CIS}.

Also, as clearly discussed by Villani in \cite{Vil},  apart from collisions, other physically relevant problem in kinetic theory of granular gases
lead to the addition of various terms which either model external physical forces, or arise from particular situations. One of these situations is
described by equation \eqref{hb}. A second one is obtained by subtracting a drift term to the Boltzmann collision operator. This leads to the
problem of finding steady states of the equation
 \begin{equation}\label{drift}
 Q(f,f) - \sigma^2 \nabla\cdot(v f) = 0.
 \end{equation}
Let us remark that in one dimension of the velocity space, equation \eqref{drift} is a particular case of equation \eqref{staztoscanibisispiga} with $\chi=-1$, which
has been solved in the previous Sub-section.

\section{Probabilistic representation of the solutions}
The core of the proofs of our results is a suitable probabilistic representation
of the solution $\mu_t$.
The idea to represent the solutions of the Kac equation  in a probabilistic way dates back, at least, to
the work of McKean \cite{McKean1966},
but it has been fully formalized and employed in the derivation of analytic results for the Kac equation
only in the last decade,
starting from \cite{CarlenCarvalhoGabetta2000} and \cite{GabettaRegazziniCLT}.

Our approach here follows the same steps used in \cite{BaLa}
 and \cite{BaLaMa} and it is based on the concept of
random recursive binary trees.
It is worth recalling that a binary tree is
a (planar and rooted) tree
where each node is either a leaf (that is, it
has no successor) or it has $2$ successors.
We define the size of the binary tree $\tau$, in symbol $|\tau|$, by the
number of internal nodes. Hence, any binary tree with $2k+1$ nodes
has size $k$ and possesses $k+1$ leaves.
Any binary tree
can be seen as a subset of
\[
\U=\{\emptyset\}\cup [\cup_{k \geq 1 } \{1,2\}^k].
\]
As usual $\emptyset$ is the root and  if $\s=(\s_1,\dots,\s_k)$
($\s_i \in \{1,2\})$ is a node of a binary tree then the length of
$\s$ is $|\s|:=k$. Moreover
$(\s,\s_{k+1}):=(\s_1,\dots,\s_k,\s_{k+1})$ and for every $1 \leq
i\leq k$, $\s|i:=(\s_1,\dots,\s_i)$ and $\s|0=\emptyset$.

We now describe a  tree evolution process which gives rise to the so
called ``random binary recursive tree''. The evolution process
starts with $T_0$, an empty tree, with just an external node (the
root). The first step in the growth process is to replace this
external node by an internal one with $2$ successors that are leave.
In this way one obtains $T_1$. Then with probability 1/2 (i.e. one
over the number of leaves) one of these $2$ leaves is selected and
again replaced by an internal node with $2$ successors. One
continues along the same rules. At every time $k$,  a binary tree
$T_k$  with $k$ internal nodes is obtained.
For more details on binary recursive trees see, for instance, \cite{drmota}.


In the rest of the paper, given a binary tree $\tau$, we shall
denote by $\CL(\tau)$ the set of the leaves of $\tau$ and by
$\CI(\tau)$ the set of the internal nodes of $\tau$.

The Wild series expansion \eqref{Wild1}-\eqref{Wild2} can be translated in
 a probabilistic representation of the solutions as sums of random variables indexed by binary recursive random trees. On a sufficiently large probability space $(\Omega,\CF,P)$ let
the following be given:
\begin{itemize}
\item a family $(X_v)_{v \in \U}$ of independent  random variables
with common  probability distribution $\bar \mu_0$;
\item a family $\big(A_0{(v)},A_1(v),A_2{(v)} \big)_{v \in \U}$ of independent positive random
vectors with the same distribution of  $(A_0,A_1,A_2)$;
\item a sequence of binary recursive random trees $(T_n)_{n\in\setN}$;
\item a stochastic process $(\nu_t)_{t\geq 0}$ with values in $\N_0$ such that
 $P\{\nu_t=k\}=e^{-t}(1-e^{-t})^{k}$ for every integer $k \geq 0$.
\end{itemize}
Write $A{(v)}= (A_0{(v)},A_1(v),A_2{(v)})$ and
assume further that
\[ (A{(v)})_{v \in \U },
\quad (T_n)_{n \geq 1}, \quad
 (X_v)_{v \in \U}
\quad\mbox{and} \quad (\nu_t)_{ t>0 } \]
are stochastically independent.

For each node $v=(v_1,\dots,v_k)$ in $\U$  set
\[
\be(v):=\prod_{i=0}^{|v|-1} A_{v_{i+1}}(v|i)
\]
and $\be(\emptyset)=1$.
Define
\[
W_0:=X_\emptyset \;\;\mbox{and} \;\; \Gamma_0:=0
\]
and, for any $n \geq 1$,
\[
W_{n}:=\sum_{v \in \CL(T_{n})} \be(v) X_v, \qquad
\Gamma_n:=\sum_{v \in \CIn(T_{n})} \be(v) A_{0}(v), \qquad
W^*_n:=W_n+\Gamma_n.
\]

\begin{proposition}\label{Prop:probint}
Equation \eqref{eq.boltzivp} has a unique solution $\phi$, which
  coincides with the characteristic function of $W^*_{\nu_t}$, i.e.
  \begin{align*}
    \phi(t,\xi) = \E[e^{i \xi W^*_{\nu_t}} ] = \sum_{n=0}^\infty e^{-t}(1-e^{-t})^{n} \E[e^{i \xi W^*_{n}} ]
    \qquad (t >0, \, \xi \in \RE).
  \end{align*}
\end{proposition}

\begin{proof}
We need some preliminary results on recursive binary trees.
A very important issue is that any binary tree
 has a recursive structure. More precisely we can use the following
recursive definition of binary trees: a binary tree $\t$ is either just an external node or an internal node
with $2$ subtrees, $\t^{(1)}, \t^{(2)}$, that are again binary trees.
For every $k\geq 0$ let $\T_{k}$ denote the set of all binary
trees with size $k$.
By Proposition 3.1 in \cite{BaLa}, we know that if
  $(T_k)_{k \geq 0}$ is a sequence of random binary recursive trees, then
for every $k \geq 1$, $j=0,\dots,k-1$ and  every $\t$ in $\T_{k}$,
\begin{equation}\label{fact-tree}
\begin{split}
P\Big \{ T^{(1)}_k=\t^{(1)}, & T^{(2)}_k=\t^{(2)}\Big| |T^{(1)}_k|=j \Big\} \\
& =P\{ T_j =\t^{(1)}\} P\{ T_{k-j-1} =\t^{(2)}\} \J\{|\t^{(1)}|=j\} \\
\end{split}
\end{equation}
and for $k\geq 1$
\begin{equation}\label{fact-tree2}
P\{|T^{(1)}_k|= j\}=\frac{1}{k}
\end{equation}
for every $j=0,\dots,k-1$.
Now observe that, in order to prove the proposition we need only to prove that
$q_n(\xi)=\E[e^{i\xi W^*_n}]$, for every $n\geq 0$. This is clearly true for $n=0$.
For $n\geq 1$, write
\[
W^*_n= A_0(\emptyset)+ \sum_{j=1}^{2} A_j{(\emptyset)} \Big\{ \Big [ \sum_{v \in \CL(T_n^{(j)})}
 \prod_{i=0}^{|v|-1} A^{(j)}_{v_{i+1}}(v|i) X^{(j)}_v \Big]
+\Big [ \sum_{v \in \CI(T_n^{(j)})}
 \prod_{i=0}^{|v|-1} A^{(j)}_{v_{i+1}}(v|i) A_0^{(j)}(v) \Big]
\Big\}
\]
where $A^{(j)}(v)=A((j,v))$, $ X^{(j)}_v=X_{(j,v)}$ and, by convention, if $\CL(T^j_n)=\emptyset$
the terms between square brackets is equal to $X^j_{\emptyset}=X_{j}$. Since $(A^{(j)}(v), X^{(j)}_v)_{v\in \U}$, $j=1,2$, are
independent, with the same distribution of $(A(v), X_v)_{v\in \U}$, using \eqref{fact-tree} and the
induction hypothesis one proves  that
\begin{equation}\label{fact2}
\E\left [e^{i\xi W^*_n}\Big |A{(\emptyset)},|T_n^{(1)}|, |T_n^{(2)}|\right ]=
\prod_{j=1}^2 q_{|T_n^{(j)}|}(\xi A_j{(\emptyset)})e^{i\xi  A_0(\emptyset)}.
\end{equation}
At this stage the conclusion  follows easily by  using \eqref{fact-tree2};
indeed:
\[
\E [e^{i\xi W^*_n}]=\E\Big[\prod_{j=1}^2 q_{|T_n^{(j)}|}(\xi A_j{(\emptyset)})
e^{i\xi  A_0(\emptyset)}\Big ]
= \frac{1}{n}\sum_{j=0}^{n-1}\E\Big[q_{j}(\xi A_1)q_{n-j-1}(\xi A_2)e^{i\xi A_0}\Big]
=q_n(\xi).
\]
 \end{proof}

\section{Proofs of Section \ref{main}}

Proposition  \ref{lemma1}, for $\gamma=2$, is proved in \cite{rosler},
while for general $\gamma \in (0,2)$
it can be seen as a special case of a (more general) result
contained in \cite{ruschendorf}. The proof of Proposition  \ref{lemma1}
given in \cite{ruschendorf} is based on some contraction properties of
the Wasserstein metrics.
Here we provide  a proof for $\gamma \in (0,2]$
based on a martingale method inspired by \cite{rosler}.
In this way we obtain some additional information
on the solution, used to prove
Proposition \ref{momenti_beta}.

In the following we need to consider a sequence $(T_n^*)_{n \geq 0}$
of (deterministic) binary trees. Any such tree can be seen as a subset of $\U$: starting from
$T_0^*=\emptyset$, for each $n$ denote by $T_n^*$ the binary tree obtained from $T_{n-1}^*$
replacing each leaf by an internal node with 2 successors. Recall that $\CL(T_n^*)$
($\CI(T_n^*)$, respectively) denotes  the set of the leaves (the internal nodes, respectively)
of $T_n^*$.

Define $\TT(\mu):=Q^+(\mu,\mu)$ and $\TT^n(\delta_m)$ as the $n$-iterate of the transfomration
 $\TT$ applied to the mass probability concentrated on the real
value $m$. Finally set
\[
M_n^*:= m \sum_{v \in \CL(T_n^*)} \be(v) + \sum_{v \in \CI(T_n^*)} \be(v) A_0(v).
\]

In the rest of the paper  $L^\gamma$ will stand for $L^\gamma(\Omega,\CF,P)$.

\begin{lemma}\label{alberi-pari1}
Let $q(\gamma)<1$ for some $\gamma$ in $(1,2]$ and $\E|A_0|^\gamma<+\infty$. Assume either
\begin{itemize}
\item[(i)] $\E(A_1+A_2)\not=1$ and $m=\E(A_0)/(1-\E(A_1+A_2))=\bar m$ or
\item[(ii)] $\E(A_1+A_2)=1$, $\E(A_0)=0$ and $m$ arbitrary.
\end{itemize}
Then
\begin{itemize}
\item[(a)] $\TT^n(\delta_m)$ is the law of $M_n^*$;
\item[(b)] $(M_n^*)_{n \geq 0}$ is a martingale with respect to
$(\CG_n^*)_{n \geq 1}$, with $\CG_n^*=\sigma(A(v): v \in T^*_{n-1})$, such that
$\E(M^*_n)=m$ for every $n$;
\item[(c)] $\sup_n \E|M^*_n|^\gamma <+\infty$, hence $(M_n^*)_{n \geq 0}$
converges a.s. and in $L^1$ to a random variable $M_\infty^*$
such that $\E(M^*_\infty)=m$ and $\E|M^*_\infty|^\gamma <+\infty$;
\item[(d)] the law  $\mu_\infty$ of $M_\infty^*$ is a solution of \eqref{stazeqVA}
in $\CM_\gamma$;
\item[(e)] If {\rm(ii)} holds true then
\[
M_\infty^*=m Z_\infty+ \sum_{n \geq 0} \sum_{v \in \CL(T^*_n)} \be(v)A_0(v)
\]
where $Z_\infty$ is the almost sure limit of $\sum_{v \in \CL(T^*_n)} \be(v)$ for $n \to +\infty$.
\end{itemize}
\end{lemma}

 \begin{proof}
 (a) is immediate for $n=1$. In fact
 \[
 \TT^1(\delta_m)=\CD(A_1(\emptyset)m+A_2(\emptyset)m+A_0(\emptyset))
 =\CD\Big(\sum_{v \in \CL(T_1^*)}\be(v)m+ \sum_{v \in \CI(T_1^*)}\be(v)A_0(v) \Big ),
 \]
 where, for every random variable $X$,  $\CD(X)$ denotes the law of $X$.
 Now, by induction, we obtain
 \[
 \begin{split}
  \TT^n(\delta_m) & =\TT(\TT^{n-1}(\delta_m)) \\
  &= \TT\Big ( \CD\Big(\sum_{v \in \CL(T_{n-1}^*)}\be(v)m+
   \sum_{v \in \CI(T_{n-1}^*)}\be(v)A_0(v) \Big ) \Big ).
   \end{split}
 \]
At this stage, denote by $T_n^{*1}$ ($T_n^{*2}$, respectively)
the left (right, respectively) binary subtree of $T_n^*$.
For every $v$ in $\U$ and $i=1,2$ set
$\be^i(v)=\be((i,v))/A_i(\emptyset)$
if $A_i(\emptyset)\not=0$ and $\be^i(v)=0$ if $A_i(\emptyset) =0$.
It is plain to check that
\begin{equation}\label{step1}
\begin{split}
&\sum_{v \in \CL(T_{n}^{*i})}\be^i(v)m +
   \sum_{v \in \CI(T_{n}^{*i})}\be^i(v)A_0((i,v)) \\
 &\text{$i=1,2$ are independent random variables with the same law of $M_{n-1}^*$.}
   \\
   \end{split}
\end{equation}
Hence,
\begin{equation}\label{step2}
\begin{split}
\TT^n(\delta_m)
&=\CD\Big (A_1(\emptyset)\big (\sum_{v \in \CL(T_{n}^{*1})}\be^1(v)m +
   \sum_{v \in \CI(T_{n}^{*1})}\be^1(v)A_0((1,v)) \big)  \\
 & \qquad + A_2(\emptyset)\big(\sum_{v \in \CL(T_{n}^{*2})}\be^2(v)m +
   \sum_{v \in \CI(T_{n}^{*2})}\be^2(v)A_0((2,v))\big) +A_0(\emptyset) \big  )  \Big )\\
   & \qquad =\CD(M_n^*).
   \\
   \end{split}
\end{equation}
As for (b) is concerned, clearly $M_n^*$ is
integrable and $\CG_n^*$ measurable. Moreover,
\[
M_n^*=M_{n-1}^*+ \sum_{v \in \CL(T_{n-1}^*)}\be(v)
[(A_1(v)+A_2(v)-1)m+A_0(v)]
\]
and hence
\[
\E[M_n^*|\CG_{n-1}^*]=M_{n-1}^*
\]
in both cases (i) and (ii). Furthermore,
$\E[M_n^*]=m$ for every $n$.
Since $M_n^*$ is a martingale and $1< \gamma \leq 2$, we can apply the
Topchii-Vatutin inequality, see e.g. \cite{alsmeyerrosler1},
to get
\[
\begin{split}
\E[|M_n^*|^\gamma] & \leq
\E[|M_0^*|^\gamma] +2 \sum_{j=1}^n \E[|M_j^*-M_{j-1}^*|^\gamma] \\
&=  m^\gamma+2 \sum_{j=1}^n \E\Big [\Big|\sum_{v \in \CL(T_{j-1}^*)}
\be(v)[(A_1(v)+A_2(v)-1)m+A_0(v)]\Big|^\gamma\Big ].
\\
\end{split}
\]
Now, since $\E[(A_1(v)+A_2(v)-1)m+A_0(v)|\CG_{n-1}^*]=\E[(A_1(v)+A_2(v)-1)m+A_0(v)]=0$,
by the Bhaar-Esseen inequality (see \cite{Bhaar-Esseen}) we obtain
\[
\E[|M_n^*|^\gamma] \leq m^\gamma + K \sum_{j=1}^n \E[\sum_{v \in \CL(T_{j-1}^*)} |\be(v)|^\gamma ]
\]
where $K=4\E[|(A_1+A_2-1)m+A_0|^\gamma]<+\infty$ by assumption.
Now it is easy to see that, for every $k \geq 0$,
\[
\E[\sum_{v \in \CL(T_{k}^*)} |\be(v)|^\gamma ]=q(\gamma)^k
\]
with $q(\gamma)<1$. Hence, $\sup_n \E|M_n^*|^\gamma<+\infty$ and,
from the elementary martingale theory, it follows that $(M_n^*)_{n \geq 0}$
converges a.s. and in $L^1$ to a random variable $M_\infty^*$
such that $\E[M_\infty^*]=m$ and $\E[|M_\infty^*|^\gamma]<+\infty$.
The proof of (c) is completed. In order to prove (d)
set $\phi_n(\xi)=\E[\exp(i\xi M_n^*)]$. By \eqref{step1}, it is clear that \eqref{step2}
is equivalent to
\[
\phi_n(\xi)=\E[\phi_{n-1}(A_1 \xi)\phi_{n-1}(\xi A_2)e^{i\xi A_0}].
\]
From (c) we know that $\phi_n(\xi)$ converges to $\phi_\infty(\xi)=\E[\exp(i\xi M_\infty^*)]$
as $n \to +\infty$. Hence, by  dominated convergence theorem, we get
\[
\phi_\infty(\xi)=\E[\phi_{\infty}(A_1 \xi)\phi_{\infty}(\xi A_2)e^{i\xi A_0}]
\]
and the proof of (d) is completed. Arguing as in the proof of
(c) it is easy to see that under (ii), the terms
$\sum_{v \in \CL(T_{n}^*)} \be(v) m$ and
$\sum_{v \in \CI(T_{n}^*)} \be(v) A_0(v)$, which
form $M_n^*$, are both uniformly integrable martingales. Hence (e) follows easily.

 \end{proof}

\begin{lemma}\label{alberi-pari2}
Let $q(\gamma)<1$ for some $\gamma$ in $(0,1]$ and $\E|A_0|^\gamma<+\infty$.
Then, for every $m$,
\begin{itemize}
\item[(a)] $\TT^n(\delta_m)$ is the law of $M_n^*$;
\item[(b)] $\sum_{v \in \CL(T^*_n)} \be(v)$  converges to $0$ in $L^\gamma$;
\item[(c)] $\Gamma_n^*=\sum_{v \in \CI(T^*_n)} \be(v)A_0(v)$
is a Cauchy sequence in $L^\gamma$ and hence it converges in $L^\gamma$ to the random variable
\[
\Gamma_\infty^*=\sum_{n \geq 0}\sum_{v \in \CL(T^*_n)} \be(v)A_0(v);
\]
\item[(d)] $M_n^*$ converges to $\Gamma_\infty^*$ in $L^\gamma$ and the law $\mu_\infty$
of $\Gamma_\infty^*$ is a solution of \eqref{stazeqVA}
in $\CM_\gamma$;
\end{itemize}
\end{lemma}

 \begin{proof}

The proof of (a) is the same of the proof of  (a) in Proposition \ref{alberi-pari1}. Furthermore, since
$\gamma  \in (0,1]$,
\[
\E\left[\big|\sum_{v \in \CL(T^*_n)} \be(v)\big|^\gamma \right ]\leq  \E\left[\sum_{v \in \CL(T^*_n)} |\be(v)|^\gamma
\right]=
 q(\gamma)^n.
\]
Using the fact that $q(\gamma)<1$, (b) follows. In order to prove (c) observe that, if $n>m$,
\[
\E[ |\Gamma_n^*-\Gamma_m^*|^{\gamma}]= \E\left[ \big|\sum_{j=m}^{n-1} \sum_{v \in \CL(T^*_n)} \be(v)A_0(v) \big|^{\gamma}
\right]
\leq \E[ |A_0|^{\gamma}] \sum_{j=m+1}^{n} q(\gamma)^j \rightarrow 0
\]
for $n, m \rightarrow +\infty$, i.e. $(\Gamma_n^*)_n$ is a Cauchy sequence in  $L^{\gamma}$. As
for assertion (d) is concerned, combining (b) and (c) we obtain that $(M_n^*)_n$ converges in $L^{\gamma}$
to $\Gamma_\infty^*$. Finally, arguing as in the proof of (d) of Proposition \ref{alberi-pari1} we obtain that
$\Gamma_\infty^*$ is a solution of \eqref{stazeqVA} in $\CM_\gamma$.
 \end{proof}

\begin{proof}[Proof of Proposition \ref{lemma1}]
The existence of a solution $\mu_{\infty}$ of \eqref{stazeqVA} in $\CM_\gamma$, as required in (a),
(b) and (c), is a consequence of Lemma \ref{alberi-pari1} and Lemma \ref{alberi-pari2}. Let us prove the
uniqueness. Let $\mu_1$ and $\mu_2$ be two solutions of \eqref{stazeqVA} in $\CM_\gamma$. Let $(Y^{(1)}_v)_v$
and $(Y^{(2)}_v)_v$ two sequences of independent random variables such that, for every  $v\in \U $, $Y^{(1)}_v$
($Y^{(2)}_v$, respectively)
has
law $\mu_1$ ($\mu_2$, respectively) and, in addition,
\[
(Y^{(1)}_v)_v, \;\;\; (Y^{(2)}_v)_v, \;\;\; (A(v))_v
\]
are stochastically independent. Then, following the same lines of (a) in Lemma \ref{alberi-pari1} and
Lemma \ref{alberi-pari2}, it is easy to see that
\[
\sum_{v \in \CL(T^*_n)} \be(v) Y^{(i)}_v + \sum_{v \in \CI(T^*_n)} \be(v)A_0(v)
\]
has law $\mu_i$ ($i=1,2$). As a consequence, if $\gamma \in (0,1]$, then
\[
\begin{split}
l_{\gamma}(\mu_1,\mu_2) & \leq \E \Big [ \big| \sum_{v \in \CL(T^*_n)} \be(v) (Y^{(1)}_v- Y^{(2)}_v) \big|^{\gamma}  \Big ]\\
 & \leq \E[ |Y^{(1)}_{\emptyset} - Y^{(2)}_{\emptyset} |^{\gamma}] \E[ \sum_{v \in \CL(T^*_n)} |\be(v)|^{\gamma} ]  \\
  & =\E[ |Y^{(1)}_{\emptyset} - Y^{(2)}_{\emptyset} |^{\gamma}] q(\gamma)^n
\end{split}
\]
and $q(\gamma)^n \rightarrow 0$ for $n \rightarrow +\infty$. Hence $\mu_1=\mu_2$ and this proves (a). As far as (b)
is concerned notice that if
$\gamma \in (1,2]$ and $\E[A_1+A_2] \neq 1$, it follows that
\[
\E[Y^{(1)}_v]= \E[Y^{(2)}_v] =\bar{m}
\]
and applying the Bhaar-Esseen  inequality 
\[
\begin{split}
l_{\gamma}^{\gamma} (\mu_1,\mu_2) & \leq   \E  \Big [ \big| \sum_{v \in \CL(T^*_n)} \be(v) (Y^{(1)}_v- Y^{(2)}_v)
\big|^{\gamma} \Big ]\\
& \leq 2 \E  [ |Y^{(1)}_{\emptyset} - Y^{(2)}_{\emptyset} |^{\gamma}] \E\Big[ \sum_{v \in \CL(T^*_n)} |\be(v)|^{\gamma}\Big ]  \\
& =2 \E[ |Y^{(1)}_{\emptyset} - Y^{(2)}_{\emptyset} |^{\gamma}] q(\gamma)^n
\end{split}
\]
and hence $\mu_1=\mu_2$ again. The case (c) follows in an analogous way since we need to consider only $\mu_1$ and $\mu_2$
in $\CM_{\gamma,m}$ (i.e. we fix the mean).
\end{proof}

\begin{proof}[Proof of Proposition \ref{momenti_beta}]
The proof of (i) is straightforward.

The proofs of (ii) and (iii) are inspired by the proof of Theorem 5.3 in \cite{DurretLiggett}. Let us first prove (ii). Note that, since $q$ is a
convex function, for every $\lambda$ in $[0,1]$, $q(\lambda \gamma + (1-\lambda)\beta) \leq \lambda q( \gamma) + (1-\lambda)q(\beta)$, and hence
$q(s)<1$ for every $\gamma \leq s \leq \beta$. In addition $\E|A_0|^s<+\infty$ since $\E|A_0|^\beta<+\infty$. Now fix $s\leq \beta$, with $1\leq k
< s \leq k+1$, $k$ integer. Then, for $x_i\geq 0$
\begin{equation}\label{DL1}
\Big ( \sum_{i=1}^3 x_i  \Big )^s= \Big ( \sum_{i=1}^3 x_i  \Big )^{\frac{s}{k+1}(k+1)}
\leq \sum_{i=1}^3 x_i^s+ \sum c_{j_1 j_2 j_3} (x_1^{j_1} x_2^{j_2} x_3^{j_3})^{\frac{s}{k+1}}
\end{equation}
for suitable constants $c_{j_1 j_2 j_3}$ and $j_i$ are integers such that
$j_i \leq k$ and $j_1 + j_2 + j_3=k+1$.
Using \eqref{DL1} it is easy to see that
\begin{equation}\label{DL2}
\E[|Y_1A_1+Y_2A_2 +A_0|^s] \leq q(s) \E|Y|^s +c_1\E[|Y^k]^{\frac{s}{k}} +c_2
\end{equation}
if $Y,Y_1,Y_2$ are independent random variables with the same law $\nu$ and $(Y,Y_1,Y_2)$
is independent of $(A_1,A_2,A_0)$. The constants $c_1$ and $c_2$ may depend on $\beta$
but not on $\nu$. Obviously \eqref{DL2} is equivalent to
\begin{equation}\label{DL3}
\int |x|^s (\TT \nu)(dx) \leq q(s) \int |x|^s \nu(dx)  +c_1\Big [\int |x|^k \nu(dx) \Big ]^{\frac{s}{k}} +c_2
\end{equation}
Let us consider first the case in which $\gamma \in (1,2]$. Choose either
$m=\bar m$ if $\E[A_1+A_2]\not =1 $ or
$m=m_0=\E[X_0]$ if $\E[A_1+A_2]=1$ and $\E[A_0]=0$, and
take $\nu=\delta_m$. From Lemma \ref{alberi-pari1}
we know that $\TT^n \delta_m$ converges weakly to $\mu_\infty$ and that
\[
\sup_n \int |x|^\gamma (\TT^n \delta_m)(dx) <+\infty.
\]
Let us now prove that if for $k \geq 1$ and
$k<s\leq k+1$, one has
\begin{equation}\label{DL4}
\sup_n \int |x|^k (\TT^n \delta_m)(dx)<+\infty \qquad \text{and}
\qquad q(s)<1
\end{equation}
then
\begin{equation}\label{DL5}
\sup_n \int |x|^s(\TT^n \delta_m)(dx)<+\infty \qquad \text{and}
\qquad \int |x|^s \mu_\infty(dx) < +\infty.
\end{equation}
In fact, applying iteratively \eqref{DL3} starting from $\nu=\delta_m$, since
$\int |x|^s \delta_m(dx)=|m|^s<+\infty$, we obtain
\[
\int |x|^s (\TT^n\delta_m)(dx) \leq |m|^s q(s)^{n} + C \sum_{j=0}^{n-1} q(s)^j
\]
for a suitable constant $C$. Hence, since $q(s)<1$,
one gets $\sup_n \int |x|^s (\TT^n\delta_m)(dx) <+\infty$. Furthermore define
$g_M(x)=|x|^s \J_{\{|x|\leq M\}} +M^s\J_{\{|x|> M\}}$
then
\[
\begin{split}
\int |x|^s \mu_\infty(dx)& =\int \liminf_{M \to +\infty} g_M(x) \mu_\infty(dx)
\leq \liminf_{M \to +\infty} \int g_M(x) \mu_\infty(dx) \\
&  \leq \liminf_{M \to +\infty} \lim_{n \to +\infty} \int g_M(x) (\TT^n \delta_m)(dx)
\leq   \lim_{n \to +\infty} \int |x|^s (\TT^n \delta_m)(dx) < +\infty. \\
\end{split}
\]
Now, since $\gamma>1$ and $\beta >2$, then
\eqref{DL4} is true for $k=1$ and $s=2$. As a consequence we obtain
\eqref{DL5} for $s=2$. Let us iterate this procedure for $k < \bar k$
with $\bar k < \beta \leq \bar k+1$. The last step starts from the validity of
\eqref{DL4} for $k=\bar k$ and $s=\beta$ whcih
implies $\int |x|^\beta \mu_\infty(dx)<+\infty$.

If $0 <\gamma \leq 1$ then
\[
\E|A_0+A_1 Y_1 +A_2 Y_2| \leq \E|Y| q(1) + \E|A_0|.
\]
Since $q(1)<1$ and $\E|A_0|<+\infty$ we get
\[
\int |x| (\TT^n\delta_m)(dx) \leq |m| q(1)^{n} + C \sum_{j=0}^{n-1} q(1)^j
\]
and hence, thanks to Lemma \ref{alberi-pari2},
\[
\sup_n \int |x| (\TT^n \delta_m)(dx)<+\infty \qquad \text{and}
\qquad  \int |x| \mu_\infty(dx) <+\infty.
\]
At this stage \eqref{DL4} is proved for $k=1$ and $s=k+1=2$ and we can go on as in the previous case.
This proves (ii).

Let us prove (iii). If $\gamma \leq 1$, from Lemma \ref{alberi-pari2} (c)
one obtains that $\mu_\infty\{[0,+\infty) \}=1$.
Now assume that $\gamma \in (1,2]$. Since $P\{A_0 \geq 0\}=1$ and $P\{A_0\not=0\}>0$,
then $\E[A_0]\not =0$ and only case (b) of Proposition \ref{lemma1} has to be considered.
By assumption $\bar m=\int x \mu_\infty(dx)>0$ and
hence from Lemma \ref{alberi-pari1} (c)-(d), since
$M_n^*$ is positive a.s. for every $n\geq 1$, again $\mu_\infty\{[0,+\infty) \}$.

As a consequence, using the fact that $\beta \geq 1$ we can write
\[
X^\beta =^{\CD} (A_0+A_1 X_1 +A_2 X_2  )^\beta \geq  A_0^\beta +A_1^\beta  X_1^\beta  +A_2^\beta  X_2^\beta
\]
($=^{\CD}$ denotes the identity in distribution) if $X,X_1,X_2$ are independent random variables
with law $\mu_\infty$ and $(X,X_1,X_2)$ and  $(A_0,A_1,A_2)$
are stochastically independent.  Then
\[
\E [X^\beta] \geq q(\beta)\E [X_1^\beta]+  \E[A_0^\beta] > q(\beta)\E [X_1^\beta]
\]
since we are assuming that $P\{A_0 \not =0 \}>0$. Hence $q(\beta)<1$.

\end{proof}

 Let us state a useful result which is proved, with slightly different notation, in
Lemma  2 of \cite{BaLaMa} (see also Proposition 4.1 in \cite{BaLa}).

\begin{lemma}\label{lemmamomenti}
Let $\gamma>0$ such that $q(\gamma)=\E[|A_1|^\gamma+|A_2|^\gamma]<+\infty$. Then, for every $n \geq 0$,
\begin{equation}\label{pesi}
\E\Big [\sum_{v \in \CL(T_n)} |\be(v)|^\gamma  \Big]=\frac{\Gamma(q(\gamma)+n)}{\Gamma(n+1)\Gamma(q(\gamma))}
=: c_n(\gamma).
\end{equation}
\end{lemma}

\begin{proof}
Given the sequence
$(T_n)_{n \geq 1}$ of random binary recursive trees, one can define a sequence $(V_n)_{n \geq 1}$
of $\U$-valued random variables such that
\[
T_{n+1}=  T_n \cup \{ (V_n,1), (V_n,2)\}
\]
for every $n \geq 0$, where $V_0=\emptyset$ and $V_n \in \CL(T_n)$. The random variable $V_n$
corresponds to the random vertex chosen to generate $T_{n+1}$ from $T_n$. Hence, by construction,
\[
P\{ V_n=v| T_1, \dots, T_n\}
= \J\{ v \in \CL(T_n)\} \frac{1}{n+1}
\]
for every $n\geq 1$.
Since $T_0=\emptyset$ and $\be(\emptyset)=1$,
$\E[\sum_{v \in \CL(T_0)} |\be(v)|^\gamma  ]=1$ and hence
\eqref{pesi} is true for $n=0$.
For $n \geq 1$,
\[
\begin{split}
\E\Big[& \sum_{v \in \CL(T_n)} |\be(v)|^\gamma  \Big]=
\E\Big[
\sum_{v \in \CL(T_{n-1})} |\be(v)|^\gamma\Big[(|A_1(v)|^\gamma+|A_2(v)|^\gamma-1)\J\{V_{n-1}=v\}+1\Big]\Big]\\
&=\E\Big[\sum_{v \in \CL(T_{n-1})} |\be(v)|^\gamma  \Big]\Big(1+\frac{q(\gamma)-1}{n}\Big)
\\
\end{split}
\]
from the independence assumptions and since $(A_1(v),A_2(v))$ has the same law of $(A_1,A_2)$ for every
$v$. Hence, by induction,
\[
\E\Big[ \sum_{v \in \CL(T_n)} |\be(v)|^\gamma  \Big]=
\prod_{j=1}^n\Big(1+\frac{q(\gamma)-1}{j}\Big)
=\frac{\Gamma(q(\gamma)+n)}{\Gamma(n+1)\Gamma(q(\gamma))}
\]
and \eqref{pesi} is proved.

\end{proof}

In the following denote by $\z_n^*$ the law of $W_n^*$.

\begin{lemma}\label{lemma5}
Assume that, for some $\gamma$ in $(0,2]$, $\E[|X_0|^\gamma+|A_0|^\gamma]<+\infty$
and $q(\gamma)<1$.
\begin{itemize}
\item[(a)] If $0 < \gamma \leq 1$, then
\[
l_\gamma (\z_n^*,\mu_\infty)\leq  c_n(\gamma) l_\gamma(\bar \mu_0,\mu_\infty)
\]
for every $n\geq 0$, where $\mu_\infty$ is the unique solution of \eqref{stazeqVA} in $\CM_\gamma$.
Furthermore, if $\gamma=1$ and $\E(X_0)=\bar m$ then
$\E(W_n^*)=\int v \z_n^*(dv)=\bar m$ for every $n \geq 1$.
\item[(b)]If $1< \gamma \leq 2$, $\E(A_1+A_2)\not=1$ and  $\E(X_0)=\bar m$,
then
\begin{equation}\label{stima1}
l_\gamma^\gamma (\z_n^*,\mu_\infty)\leq  2 c_n(\gamma) l_\gamma^\gamma(\bar \mu_0,\mu_\infty)
\end{equation}
where $\mu_\infty$ is the unique solution of \eqref{stazeqVA} in $\CM_\gamma$.
Furthermore $\E(W_n^*)=\int v \z_n^*(dv)=\bar m$ for every $n \geq 0$.
\item[(c)] If $1 < \gamma \leq 2$, $\E(A_0)=0$, $\E(A_1+A_2) =1$,  $\E(X_0)=m_0$ {\rm(}$m_0$ arbitrary{\rm)},
and $\mu_\infty$ is the unique solution of \eqref{stazeqVA} in $\CM_{\gamma,m_0}$, then
\eqref{stima1} holds and $\E(W_n^*)=\int v \z_n^*(dv)= m_0$ for every $n \geq 0$.
\end{itemize}
\end{lemma}

\begin{proof} The existence and uniqueness of $\mu_\infty$ in the three
cases (a), (b) and (c)
is guaranteed by Proposition \ref{lemma1}. On a sufficiently large probability
space $(\Omega,\CF,P)$ consider a sequence  $(Y_v)_{v \in \U}$,
such that
\begin{itemize}
\item $(A(v))_{v \in \U}$,$(T_n)_{n \geq 0}$,
and $(X_v,Y_v)_{v \in \U}$ are independent;
\item  $(X_v,Y_v)$ are independent and identically distributed for $v$ in $\U$,
and each $(X_v,Y_v)$ is an optimal transport plan for
$l_\gamma(\bar \mu_0,\mu_\infty)$, i.e.
the law of $X_v$ is $\bar \mu_0$, the law of $Y_v$ is
$\mu_\infty$ and $\E|X_v-Y_v|^{\gamma}=l_\gamma^{\max(1,\gamma)}(\bar \mu_0,\mu_\infty)$.
\end{itemize}
Let us set
$U_n^*=\sum_{v \in \CL(T_n)}  Y_v \be(v)+\Gamma_n.$
We now show that, for every $n$, the law of $U^*_n$ is $\mu_\infty$. In point of fact
\[
\begin{split}
    \E[e^{i\xi U_n^*}]& =
    \E\Big[ \sum_{\bar v \in \CL(T_{n-1})} \J\{V_{n-1}=\bar v\}  \\
    & \exp\Big\{i\xi       \Big (    \sum_{ v \not = \bar v} \be(v) Y_v +\Gamma_{n-1}
    + \be(\bar v) (\underbrace{A_1(v) Y_{\bar v1} + A_2 (\bar v) Y_{\bar v2} + A_0(\bar v)}_{=^d Y_{\bar v}})
      \Big )  \Big\}\Big]  \\
    &= \E\Big[ \sum_{\bar v \in \CL(T_{n-1})} \J\{V_{n-1}=\bar v\} e^{i\xi U^*_{n-1}}
     \Big]
     \\
    &
     = \E[e^{i\xi U^*_{n-1}}] \\
  \end{split}
  \]
  where $V_n$ is defined as in the proof of Lemma \ref{lemmamomenti}.
 Hence, by induction, $U_n^*$ has the same law of  $Y_{\emptyset}$, that is $\mu_\infty$.
Now denote by $\CG$ the $\s$--field $\s(A(v) : v \in \U, (T_n)_{n \geq 1} ) $ and observe that
\[
\begin{split}
l_\gamma^{\max(1,\gamma)}(\z_n^*,\mu_\infty) & \leq \E|W^*_n-U_n^*|^\gamma
=\E\Big [\E \Big  [|\sum_{v \in \CL(T_n)} \be(v)(X_v-Y_v) |^\gamma\Big |\CG\Big  ]\Big ] \\
&  \leq  k_\gamma  \E\Big [ \sum_{v \in \CL(T_n)} |\be(v)|^\gamma \E\Big [|X_v-Y_v |^\gamma\Big |\CG\Big  ]\Big ] .
\\
\end{split}
\]
The last inequality is immediate for $\gamma \leq 1$ with $k_\gamma =1$ while,
if $1<\gamma \leq 2$, it follows with $k_\gamma=2$ from Bhaar-Esseen inequality, 
since $\E[X_v]=\E[Y_v]$ which  implies
$\E \Big  [\sum_{v \in \CL(T_n)} \be(v)(X_v-Y_v)\Big |\CG\Big  ]=0$.
Hence, using \eqref{pesi},
\[
l_\gamma^{\max(1,\gamma)}(\z_n^*,\mu_\infty)
\leq  k_\gamma
\E\Big [ \sum_{v \in \CL(T_n)} |\be(v)|^\gamma \Big ] l_\gamma^{\max(1,\gamma)}(\mu_0,\mu_\infty)
= k_\gamma  c_n(\gamma) l_\gamma^{\max(1,\gamma)}(\mu_0,\mu_\infty) .
\]
In order to conclude the proof, let us study $\E[W_n^*]$ when $\gamma$ belongs to $[1,2]$.
Observe that
\[
\E[W_n^*]= \E[W_{n-1}^*]
+\E\Big [\sum_{v \in \CL(T_{n-1}) } \be(v) [A_0(v) + A_1(v)X_{v1}+A_2(v)X_{v2}-X_v ]
\J\{V_n=v\}\Big ].
\]
If $\CG_{n-1}=\s( T_1,\dots,T_n; \be(v) : v \in \CL(T_{n-1})  )$, then
\[
\begin{split}
\E\Big [ & \sum_{v \in \CL(T_{n-1}) } \be(v) [A_0(v) + A_1(v)X_{v1}+A_2(v)X_{v2}-X_v ]
\J\{V_n=v\}\Big ] \\
& =\E \Big [  \sum_{v \in \CL(T_{n-1}) } \be(v) \J\{V_n=v\} \E \big [A_0(v) + A_1(v)X_{v1}+A_2(v)X_{v2}-X_v
\big | \CG_{n-1} \big ]\Big ]
\\
\end{split}
\]
and
\[
\E \big [A_0(v) + A_1(v)X_{v1}+A_2(v)X_{v2}-X_v
\big | \CG_{n-1} \big ]=\E[A_1+A_2-1]\E[X_0]+\E[A_0]=0
\]
either in case (b) and (c) or in case (a) when $\gamma=1$ and
$\E[X_0]=\bar m$.
Hence, $\E[W^*_n]=\E[W^*_0]=\E[X_0]$, which completes the proof.
\end{proof}

\begin{proof}[Proof of Proposition \ref{Pr1-Bis}]
Using Proposition \ref{Prop:probint},
to the convexity of the  Wasserstein
distance $l_\gamma$ $(\gamma<1)$ and Lemma \ref{lemma5} (a) one gets
\[
\begin{split}
&l_\gamma(\mu_t,\mu_\infty) \leq \sum_{n\geq 0} e^{-t}(1-e^{-t})^{n} l_\gamma(\z_{n}^*,\mu_\infty)
\\
& \leq \sum_{n \geq 0} e^{-t}(1-e^{-t})^{n}
c_n(\gamma) l_\gamma(\bar \mu_0,\mu_\infty)
= e^{-t(1-q(\gamma))} l_\gamma(\bar \mu_0,\mu_\infty).\\
\end{split}
\]
\end{proof}

\begin{proof}[Proof of Proposition \ref{Pr1}]
Since $q(1)<1$, in this case
$\E[A_1+A_2]\not=1$. Hence, thanks to Proposition \ref{Prop:probint},
to \eqref{pesi} and to the convexity of the  Wasserstein
distance
\[
\begin{split}
&l_1(\mu_t,\mu_\infty) \leq \sum_{n\geq 0} e^{-t}(1-e^{-t})^{n} l_1(\z_{n}^*,\mu_\infty)
\\
& \leq \sum_{n \geq 0} e^{-t}(1-e^{-t})^{n}
c_n(1) l_1(\mu_0,\mu_\infty)
= e^{-t(1-q(1))} l_1(\mu_0,\mu_\infty).\\
\end{split}
\]
Furthermore, if $\E[X_0]=\bar m$, then
\[
\int v \mu_t(dv) =\sum_{n\geq 0} e^{-t}(1-e^{-t})^{n}  \E[W_n^*]=
\sum_{n\geq 0} e^{-t}(1-e^{-t})^{n}\bar m =\bar m,
\]
since
 $ \E[W_n^*]= \bar m$ as stated in (b)
of Lemma \ref{lemma5}.
\end{proof}

\begin{proof}[Proof of Proposition \ref{Pr2}]
Using Proposition \ref{Prop:probint},
to the convexity of t
 $l_\gamma^\gamma$ $(\gamma\geq 1)$ and Lemma \ref{lemma5} (b) one gets
\[
\begin{split}
&l_\gamma^\gamma(\mu_t,\mu_\infty) \leq \sum_{n\geq 0} e^{-t}(1-e^{-t})^{n} l_\gamma^\gamma(\z_{n}^*,\mu_\infty)
\\
& \leq \sum_{n \geq 0} e^{-t}(1-e^{-t})^{n}
c_n(\gamma) l_\gamma^\gamma(\bar \mu_0,\mu_\infty)
= e^{-t(1-q(\gamma))} l_\gamma^\gamma(\bar \mu_0,\mu_\infty).\\
\end{split}
\]
Arguing exactly in as in the previous proof using Lemma \ref{lemma5} (b)
in place of Lemma \ref{lemma5} (a) one proves
$\int v \mu_t(dv)=\bar m$.
\end{proof}

\begin{proof}[Proof of Proposition \ref{Pr3}]
Analogous to the proof of
Proposition \ref{Pr2}, using Lemma \ref{lemma5} (c)
in place of Lemma \ref{lemma5} (b).
\end{proof}


\begin{thebibliography}{31}


\bibitem{alsmeyerrosler1}
\textsc{G.~Alsmeyer} and \textsc{U.~R{\"o}sler} (2003).
The best constant in the {T}opchii-{V}atutin inequality for
  martingales.
{\em Statist. Probab. Lett.} \textbf{65} 199--206.



\bibitem{BaLa}\textsc{F. Bassetti} and \textsc{L. Ladelli} (2010)
Self similar solutions in one-dimensional kinetic models:
a probabilistic view.   arXiv:1003.5527

\bibitem{BaLaMa}
\textsc{F.~Bassetti, L.~Ladelli} and \textsc{D.~Matthes} (2010).
Central limit theorem for a class of one-dimensional kinetic
  equations. {\em  Probab. Theory Related Fields}
DOI:10.1007/s00440-010-0269-8 (Published on line) .


\bibitem{BaLaRe}
 \textsc{ F.~Bassetti, L.~Ladelli} and \textsc{E.~Regazzini} (2008)
  {Probabilistic study of the speed of approach to equilibrium for an inelastic Kac model.}
  \textit{J. Stat. Phys.} \textbf{133} 683--710.

\bibitem{BaTo}
\textsc{F. Bassetti} and \textsc{G. Toscani} (2010)
 Explicit equilibria in a kinetic model of gambling. \emph{Phys. Rev. E}  \textbf{81},
066115.

\bibitem{bisispigatoscani} \textsc{M. Bisi, G. Spiga} and  \textsc{G. Toscani} (2009).
Kinetic models of conservative economies with wealth redistribution. \textit{Commun. Math. Sci.} \textbf{ 7}  901--916.


\bibitem{Bob88}
 \textsc{ A.V. Bobylev} (1988).
  The theory of the spatially Uniform Boltzmann equation for Maxwell molecules.
  {\em Sov. Sci. Review C} \textbf{7},  112--229.

\bibitem{BoCe}\textsc{A.V. Bobylev} and \textsc{C. Cercignani} (2003).
Self-similar asymptotics for the Boltzmann equation with inelastic and elastic interactions.
{\em J. Statist. Phys.} \textbf{110} 333--375.


\bibitem{CeGaBo}
\textsc{A.~V. Bobylev, C.~Cercignani} and \textsc{I.~M. Gamba} (2009).
\newblock On the self-similar asymptotics for generalized nonlinear kinetic
  maxwell models.
\newblock {\em Comm. Math. Phys.} \textbf{291} 599--644.






\bibitem{CarlenCarvalhoGabetta2000}
\textsc{E.A. Carlen, M.C. Carvalho} and \textsc{E .Gabetta} (2000).
  {Central limit theorem for {M}axwellian molecules and truncation of the {W}ild expansion.}
  \textit{Comm. Pure Appl. Math.} \textbf{ 53} 370--397 .

\bibitem{CarrCordTosc}
 \textsc{J.A. Carrillo, S. Cordier} and \textsc{G. Toscani} (2009)
  {Over-populated tails for conservative-in-the-mean inelastic Maxwell models.}
  {\it Discrete Contin. Dyn. Syst.} {\bf 24} no. 1, 59--81.

\bibitem{Cer}
\textsc{C. Cercignani} (1988). \emph{The Boltzmann equation and its applications},
\newblock {Springer Series in Applied Mathematical Sciences},
  Vol.\textbf{67} Springer--Verlag 1988.

\bibitem{Cer94}
\textsc{C. Cercignani, R. Illner} and \textsc{M. Pulvirenti} (1994).
\newblock The mathematical theory of dilute gases,
\newblock {\em Springer Series in Applied Mathematical Sciences},
  Vol.\textbf{  106} Springer--Verlag 1994.

\bibitem{CIS}
\textsc{C. Cercignani, R. Illner} and \textsc{C. Stoica} (2001)
 On diffusive equilibria in generalized kinetic theory.
\textit{J. Stat. Phys.} \textbf{105} 337--352.

\bibitem{ChaCha00}
\textsc{A.~Chakraborti} and  \textsc{B.K.~Chakrabarti} (2000) Statistical
  Mechanics of Money: Effects of Saving Propensity, {\em Eur. Phys. J. B}
  \textbf{17}, 167-170.

\bibitem{review}
\textsc{A.\ Chatterjee} and \textsc{B.K.\ Chakrabarti} (2007).
   Kinetic Exchange Models for Income and Wealth Distributions.
  \emph{Eur.~Phys.~J.~B\/} {\bf 60} , 135--149.

\bibitem{book1}
\textsc{A.\ Chatterjee, Y.\ Sudhakar}  and \textsc{B.K.\ Chakrabarti} (2005).
  {\it Econophysics of Wealth Distributions}
   New Economic Windows Series, Springer, Milan, 2005.

\bibitem{CordPareTosc}
 \textsc{S. Cordier, L. Pareschi} and \textsc{G. Toscani} (2005)
  {On a kinetic model for a simple market economy.}
  {\it J. Stat. Phys.} {\bf 120} no. 1-2, 253--277.



\bibitem{drmota}
\textsc{M.~Drmota} (2009). \newblock {\em An interplay between combinatorics and probability.}
\newblock {\em Random trees}.
\newblock Springer Wien NewYork, Vienna.


\bibitem{DurretLiggett}
\textsc{R. Durrett} and \textsc{T. Liggett} (1983).
 Fixed points of the smoothing transformation.
 \textit{Z. Wahrsch. Verw. Gebiete}  {\bf 64}  275--301.

\bibitem{GabettaRegazziniCLT}
 \textsc{ Gabetta E.} and \textsc{Regazzini E.} (2008).
  {Central limit theorem for the solution of the Kac equation.}
    Ann. Appl. Probab. {\bf 18}, 2320-2336.

\bibitem{Kac}
\textsc{M. Kac} (1959)
  {\it Probability and related topics in physical sciences}
  (Lectures in Applied Mathematics.
  Proceedings of the Summer Seminar, Boulder, Colo., 1957, Vol. I
  Interscience Publishers, London-New York 1959.)


\bibitem{MatthesToscani}
  \textsc{Matthes, D.} and \textsc{Toscani, G.} (2008).
  On steady distributions of kinetic models of conservative economies.
  \textit{ J. Statist. Phys.} {\bf 130}  1087-1117.




\bibitem{McKean1966}
\textsc{ H.~P  McKean Jr } (1966).
  {Speed of approach to equilibrium for {K}ac's caricature of a {M}axwellian gas.}
  \textit{Arch. Rational Mech. Anal.} \textbf{21} 343--367.


\bibitem{PulvirentiToscani}  \textsc{ Pulvirenti, A.}  and  \textsc{Toscani, G.} (2004).
  {Asymptotic properties of the inelastic {K}ac model.}
  \textit{J. Statist. Phys.} {\bf 114} 1453--1480.


\bibitem{RachevRuschendorf}
 \textsc{Rachev, S. T.} and  \textsc{R\"uschendorf, L.} (1995).
Probability metrics and recursive algorithms.  \textit{Adv. Appl. Prob.} {\bf 27}, 770--799.


\bibitem{rosler} \textsc{ U. R\"osler} (1992). A fixed point theorem for distributions.  Stochastic Process. Appl. {\bf 42}   195--214.

\bibitem{ruschendorf}  \textsc{L. Rüschendorf} (2006). On stochastic recursive equations of sum and max type.  J. Appl. Probab.  {\bf 43} 687--703.





\bibitem{Vil}
\textsc{C. Villani} (2006) Mathematics of granular materials.
 \textit{J. Stat. Phys.} \textbf{124} 781--822.


\bibitem{Bhaar-Esseen}
\textsc{ B. von Bahr} and \textsc{C.G. Esseen} (1965).
  Inequalities for the $r$th absolute moment of a sum
  of random variables, $1\leq r\leq 2$.
  \textit{Ann. Math. Statist} \textbf{36}  299--303.

\bibitem{Wild1951}
\textsc{ E. Wild}  (1951).
  {On {B}oltzmann's equation in the kinetic theory of gases.}
  \textit{Proc. Cambridge Philos. Soc.} \textbf{47} 602--609.



\end{thebibliography}

\end{document}